\documentclass[conference]{IEEEtran}
\IEEEoverridecommandlockouts
\usepackage{cite}
\usepackage{amsmath,amssymb,amsfonts,bm}
\usepackage{algorithmic}
\usepackage{graphicx}
\usepackage{textcomp}
\usepackage{xcolor}
\usepackage{cases}
\usepackage[caption=false,font=normalsize,labelfont=sf,textfont=sf]{subfig}

\def\BibTeX{{\rm B\kern-.05em{\sc i\kern-.025em b}\kern-.08em
    T\kern-.1667em\lower.7ex\hbox{E}\kern-.125emX}}

\newtheorem{lemma}{Lemma}

\begin{document}

\title{Fundamental Limits of Thermal-noise Lossy Bosonic Multiple Access Channel
\thanks{This material is based upon work supported by the National Science Foundation under Grant No. CCF-2006679.}
}

\author{\IEEEauthorblockN{Evan J.D. Anderson}
\IEEEauthorblockA{\textit{Optical Sciences} \\
\textit{University of Arizona}\\
Tucson Arizona, USA \\
ejdanderson@optics.arizona.edu}
\and
\IEEEauthorblockN{Boulat A. Bash}
\IEEEauthorblockA{\textit{Electrical \& Computer Engineering,  Optical Sciences } \\
\textit{University of Arizona}\\
Tucson Arizona, USA \\
boulat@arizona.edu}
}

\maketitle

\begin{abstract}
 Bosonic channels describe quantum-mechanically many practical communication links such as optical, microwave, and radiofrequency. We investigate the maximum rates for the bosonic multiple access channel (MAC) in the presence of thermal noise added by the environment and when the transmitters utilize Gaussian state inputs. We develop an outer bound for the capacity region for the thermal-noise lossy bosonic MAC. We additionally find that the use of coherent states at the transmitters is capacity-achieving in the limits of high and low mean input photon numbers. Furthermore, we verify that coherent states are capacity-achieving for the sum rate of the channel. In the non-asymptotic regime, when a global mean photon-number constraint is imposed on the transmitters, coherent states are the optimal Gaussian state. Surprisingly however,  the use of single-mode squeezed states can increase the capacity over that afforded by coherent state encoding when each transmitter is photon number constrained individually.
\end{abstract}


\section{Introduction}

The multiple access channel (MAC) is the principal building block of many practical networks.  Quantum information \cite{nielsen00quantum,wilde16quantumit2ed} governs the fundamental limits of physical channels comprising any network, and offers substantial benefits in their performance \cite{holevo_capacity_1999,giovannetti03broadband,giovannetti_minimum_2004,giovannetti_ultimate_2014} and security \cite{scarani09rmpQKD,pirandola19QKDreview,bash15covertbosoniccomm,bullock20discretemod,gagatsos20codingcovcomm}. While the MAC has been studied extensively in classical network information theory \cite{elgamal12nit}, the quantum perspective has been underexplored.  With notable exception of \cite{yen_multiple-access_2005,yen05phdthesis}, previous work has largely focused on the quantum channels that act on finite-dimensional qudits \cite{winter01qMACcap,yard05qMACtheorems,ahlswede06qMACstrongconv}.  While recent results focused on coding \cite{Qi_2018,hayashi21cqMACcoding,hayashi21compMAC},  entanglement-assisted communication \cite{shi_computable_2021, shi2021entanglement, laurenza2017general, anshu2019near}, and secrecy \cite{aghaee2020private,aghaee2019classical}, there is a gap in understanding of the fundamental limits of the bosonic multiple access communication in the presence of thermal-noise.

While bosonic channels model quantum-mechanically many practical channels (including free-space and fiber optical, microwave, and radiofrequency (RF)), they are particularly useful in optical communications. This is because noises of quantum-mechanical origin limit the performance of advanced high-sensitivity photodetection systems \cite{sinclair19nanowire,mccaughan19nanowire,lee18tes} and the bosonic MAC in particular accurately represents high-speed optical interconnects between and within silicon microchips. Furthermore, quantum methodology includes resources such as squeezed states, shared entanglement, and joint detection receivers that can substantially increase communication capacity. Indeed, the bosonic channel model allows the fundamental limits in throughput and security to be explored by lifting all the assumptions on the transceiver and the adversary except those allowed by the laws of physics. Previous work developed and analyzed the pure-loss bosonic MAC when no excess noise was injected by the environment and the transmitters were restricted to Gaussian inputs \cite{yen_multiple-access_2005,yen05phdthesis}. However, such a model does not completely describe practical communication systems as it does not take into account noise in the system. While this previous work \cite{yen_multiple-access_2005,yen05phdthesis} demonstrated that utilization of coherent states at the input and heterodyne detection are capacity-reaching in the high signal-to-noise (SNR) regime for the pure-loss MAC, it did not analyze the asymptotic limit of low mean photon-number constraints at the input. Progress has also been made on the use of entanglement-assistance in the bosonic MAC \cite{shi_computable_2021}, but the capacity region for the unassisted thermal-noise lossy bosonic MAC still remains underexplored.

We present a model that allows for thermal noise from the environment to be injected into the system. This model allows us to analyze the thermal-noise lossy bosonic MAC via development of maximum rates when Gaussian state inputs are used at the receivers. It additionally allows for the investigation of capacity bounds when asymptotically large and small mean photon number at the transmitters are employed. The inclusion of thermal noise and understanding the limit of low signal power are essential in performing covert communication analysis where an adversary is unable to distinguish between a signal from the transmitter(s) and background noise \cite{gagatsos20codingcovcomm,bash12sqrtlawisit,bash13squarerootjsacnonote,bash15covertcommmag,bash15covertbosoniccomm, bullock20discretemod, anderson21bosonicbroadast}. In evaluating the asymptotic limits of high and low mean photon number at the inputs, we find that coherent states are capacity-achieving (Lemmas \ref{lemma:1} and \ref{lemma:2}, respectively). However, in the finite mean photon-number regime we find that the use of single-mode squeezed states can be beneficial over the use of coherent states unless there is a global mean photon-number constraint at the transmitters in which case coherent states are the optimal Gaussian states (Section \ref{sec:fixed-mean-photon}).

After formally defining our channel model and stating necessary previous technical results in Section \ref{sec:preliminaries}, in Section \ref{sec:main} we build on \cite{yen_multiple-access_2005,yen05phdthesis} to develop the maximum rates for the lossy thermal-noise bosonic MAC when the transmitters are limited to Gaussian inputs. We conclude in Section \ref{sec:discussion} with discussion of the implications of our results on future work.

\section{Preliminaries}
\label{sec:preliminaries}

\subsection{Thermal-Noise Bosonic MAC} \label{sec:channelmodel}

\begin{figure} 
\centerline{\includegraphics[width=0.5\textwidth]{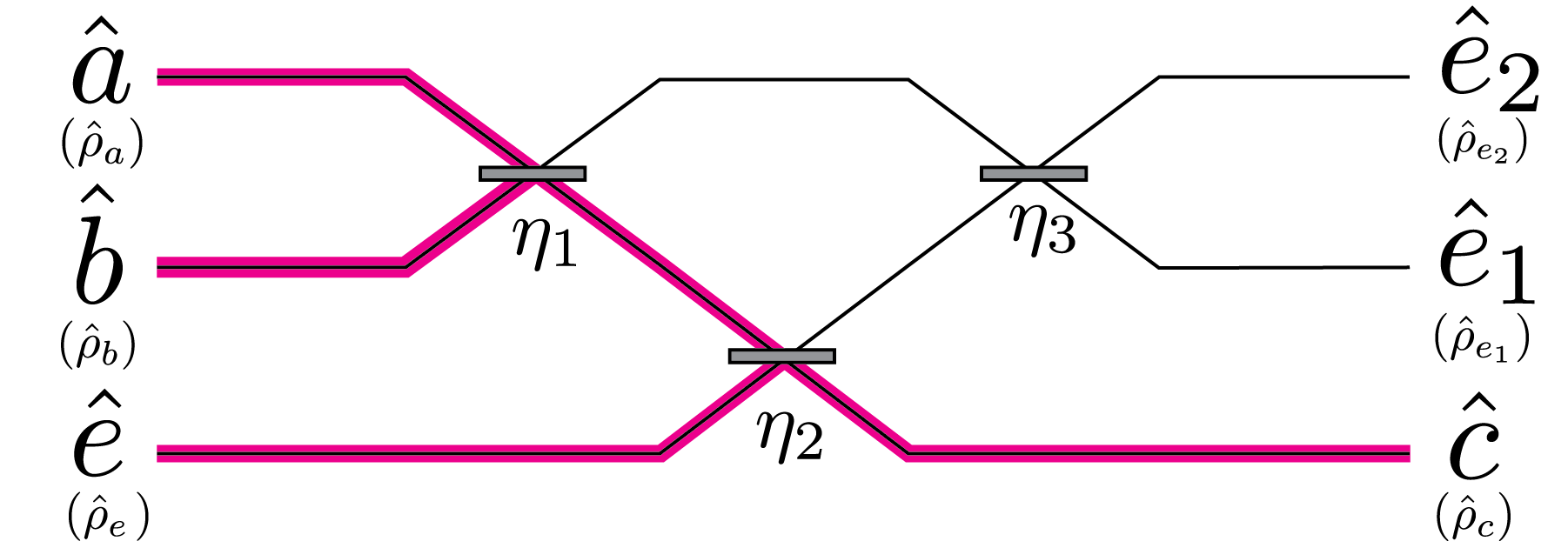} } 
\caption{The generalized three-input, three-output, lossy thermal-noise bosonic channel modeled by beamsplitters with transmissivity $\eta_1, \eta_2, \eta_3$, an assumed known phase shift set to zero for each beamsplitter, and modal input and output operators $\hat{a}, \hat{b}, \hat{e}, \hat{e}_1, \hat{e}_2, \hat{c}$. Input modal annihilation operators $\hat{a}$ and $\hat{b}$ represent transmitters Alice and Bob while the third $\hat{e}$ is the environment. The receiver, Charlie, is modeled by output modal annihilation operator $\hat{c}$ while $\hat{e}_1, \hat{e}_2$ represent photon loss to the environment. The path photons take from each of the input modes through the beamsplitters and arrive at the receiver is highlighted.}\label{fig:model}
\end{figure}

 This pure-loss channel analyzed by Yen and Shapiro in \cite{yen_multiple-access_2005, yen05phdthesis} consists of two input ports (Alice and Bob) and two output ports (Charlie and the environment) along with a single beamsplitter which modeled power coupling to Charlie. However, the applications of the lossy bosonic MAC in \cite{yen_multiple-access_2005, yen05phdthesis} are limited to systems that are not afflicted by excess thermal noise.  Including thermal noise results in a richer model.  Therefore, in this paper, we include thermal noise by adding an input port and a corresponding output port to the model, which also requires an additional two beamsplitters to model the modal annihilation operator relationships \cite{reck94beamsplitter, clements17uniinterf}. This channel is depicted in Fig.~\ref{fig:model}. Two input modes $\hat{a}, \hat{b}$ represent the transmitters Alice and Bob, while a third input mode $\hat{e}$ is the environment and can inject thermal noise. This model also requires three beamsplitters $\eta_1, \eta_2, \eta_3$ where we ignore their phase shift $e^{i\phi_j}, j= 1,2,3$, setting $\phi_j=0$ as is customary in the literature. The three output modes consist of the receiver Charlie, $\hat{c}$, and two environment modes, $\hat{e}_1, \hat{e}_2$, which can be traced out to account for photon loss to the environment. The modal relationship between the input modes and the output mode at the receiver is given by:
\begin{align}
\hat{c} = \sqrt{\eta_1\eta_2}\hat{a} +  \sqrt{(1-\eta_1)\eta_2} \hat{b} + \sqrt{1-\eta_2}\hat{e}. \label{eq:modemodel}  
\end{align}

We note that this model can be generalized to $K > 2$ transmitters. In this scenario, $K$ input ports are reserved for transmitters and one additional input is needed for the environment. Likewise, a single output port is needed for the receiver while $K$ output ports are reserved for the environment. In this generalized model, $\frac{K(K+1)}{2}$ arbitrary beamsplitters with transmissivity $\eta_j$, $j = 1,2,\dotsc, \frac{K(K+1)}{2}$ are also required \cite{reck94beamsplitter, clements17uniinterf}. Here, we focus on $K=2$ as seen in Fig.~\ref{fig:model}.

To define the capacity region for the two-transmitter bosonic MAC in Fig.~\ref{fig:model}, we must first define an achievable rate pair. A rate pair ($R_A,R_B$) is \emph{achievable} if there exists a sequence of (($2^{nR_A},2^{nR_B}), n$) codes such that the probability of error decays to zero as the length of the codes, $n$, increases to infinity. Winter developed the capacity region for the quantum MAC \cite{winter01qMACcap} requiring finite-dimensional Hilbert spaces. Nevertheless, we can use Winter's capacity region by extending it to infinite-dimension Hilbert spaces using a limiting argument \cite[Appendix A.1]{yen_multiple-access_2005}. Then the capacity region for the two-sender bosonic MAC is then defined as the convex closure of the union of all achievable rate pairs satisfying
\begin{align}
    R_{A} & \leq  \int p_{B}(\beta) S\left(\bar{\rho}_{\beta}^{B}\right) \mathrm{d} \beta - \bar{S} \\
    R_{B} & \leq \int p_{A}(\alpha) S\left(\bar{\rho}_{\alpha}^{A}\right) \mathrm{d} \alpha- \bar{S} \\
    R_A + R_B & \leq S(\bar{\rho}) - \bar{S} \label{eq:sum-rate}
\end{align}
where $\alpha, \beta \in \mathbb{C}$ are Alice and Bob's inputs with distributions $p_A(\alpha), p_B(\beta)$ and $\bar{\rho}_{\alpha}^{A}, \bar{\rho}_{\beta}^{B}, \bar{\rho}$ are the mean density operators at the receiver for Alice, Bob and over the product distribution $p_A(\alpha)p_B(\beta)$ for the state $\hat{\rho}(\alpha, \beta)$ respectively:
 \begin{align}
     \bar{\rho}_{\beta}^{B} &=\int p_{A}(\alpha) \hat{\rho}(\alpha, \beta) \mathrm{d} \alpha \\
     \bar{\rho}_{\alpha}^{A} &=\int p_{B}(\beta) \hat{\rho}(\alpha, \beta) \mathrm{d} \beta \\
     \bar{\rho} &=\iint p_{A}(\alpha) p_{B}(\beta) \hat{\rho}(\alpha, \beta) \mathrm{d} \alpha \mathrm{d} \beta.
 \end{align}
Additionally, $ \bar{S}=\iint p_{A}(\alpha) p_{B}(\beta) S(\bar{\rho}(\alpha, \beta)) \mathrm{d} \alpha \mathrm{d} \beta$.
 
While the sum-rate capacity, \eqref{eq:sum-rate}, is achieved using coherent-state encoding described in Sec.~\ref{sec:code}, only the bounds exist for the individual capacities, the exact solution is an open problem. As such, we consider Gaussian state inputs described next. 

\subsection{Gaussian-Input Bosonic MAC} \label{sec:gaussianqi}
 A single-mode quantum state $\hat{\rho}$ is called Gaussian if its Wigner function is of the form \cite{weedbrook12gaussianQIrmp}:
\begin{align}
W(\bm{\mu})=\frac{\exp \left[-\frac{1}{2}(\bm{\mu}-\bm{\bar{\mu}})^{\top} \bm{\Sigma}^{-1}(\bm{\mu}-\bm{\bar{\mu}})\right]}{2 \pi \sqrt{|\bm{\Sigma}|}},
\end{align}
where $|\mathbf{A}|$ and $\mathbf{A}^\top$ respectively denote the determinant and transpose of matrix $\mathbf{A}$. The mean (displacement) $\bm{\bar{\mu}}=[\bar{\mu}_1 ~ \bar{\mu}_2]^\top$ and covariance matrix $\bm{\Sigma}=\begin{bmatrix}\sigma^2_{11} ~ \sigma^2_{12} \\ \sigma^2_{21} ~ \sigma^2_{22}\end{bmatrix}$ fully characterize $\hat{\rho}$, and are given for quadrature components $\hat{p}=\mathrm{Im}\left(\hat{a}\right)$ and $\hat{q}=\mathrm{Re}\left(\hat{a}\right)$ of the annihilation operator $\hat{a}$.  We do not employ entanglement between quadratures, setting $\sigma^2_{12}=\sigma^2_{21}=0$. States of particular interest to us are:
\begin{itemize}
  \item vacuum states: $\bm{\bar{\mu}}=0, \sigma^2_{11}=\sigma^2_{22}=\frac{1}{4}$
  \item coherent states: $|\bm{\bar{\mu}}|>0, \sigma^2_{11}=\sigma^2_{22}=\frac{1}{4}$
  \item single-mode squeezed states: $|\bm{\bar{\mu}}|\ge0, \sigma^2_{11}=\frac{1}{4}e^{2r} , \sigma^2_{22}=\frac{1}{4}e^{-2r}$ where $r \in \mathbb{R}$ is the squeezing parameter
  \item thermal states: $\bm{\bar{\mu}}=0, \sigma^2_{11}=\sigma^2_{22}=\frac{1}{2}(\bar{n}_T + \frac{1}{2})$ where $\bar{n}_T > 0$ is the mean photon number.
\end{itemize}

The squeezed, coherent, and vacuum states are minimum uncertainty states, the product of their quadrature variances is $\frac{1}{16}$.  When squeezing in one quadrature the uncertainty in that quadrature is decreased while the other is increased at the cost of $\frac{1}{4} (e^{2r} + e^{-2r}) - \frac{1}{2} = \frac{1}{2}\operatorname{cosh}(2r) - \frac{1}{2}$ additional photons.

The generalized bosonic channel described in Section \ref{sec:channelmodel} preserves Gaussianity.  Thus, if Alice and Bob employ Gaussian input states, then Charlie's state is also Gaussian. The covariance matrix at the receiver in terms of the covariance matrices of the three inputs is given by:
\begin{align}
\bm{V} = \eta_1\eta_2\bm{X} + (1-\eta_1)\eta_2\bm{Y} + (1-\eta_2)\bm{Z}, \label{eq:covariancerelationship}
\end{align}
\noindent where
\begin{align}
&\bm{X} = \frac{1}{4}\begin{bmatrix}e^{2r_A} & 0 \\ 0 & e^{-2r_A}\end{bmatrix}, \label{eq:xyz} \quad
\bm{Y} = \frac{1}{4}\begin{bmatrix}e^{2r_B} & 0 \\ 0 & e^{-2r_B}\end{bmatrix} \\ &\bm{Z} = \frac{1}{4}\begin{bmatrix}2\bar{n}_T + 1 & 0 \\ 0 & 2\bar{n}_T + 1\end{bmatrix}, \quad \bm{V}=\begin{bmatrix}V_1& V_{12}\\V_{12}&V_2\end{bmatrix}. \notag
\end{align}
Alice and Bob's squeezing parameters are given by $r_A, r_B$ respectively, $\bm{X}, \bm{Y}, \bm{Z}$ are the covariance matrices for the input modes of the two transmitters and environment respectively, and $\bm{V}$ is the covariance matrix at Charlie's output mode $\hat{c}$. The components of $\bm{V}$ are explicitly given by:
\begin{align}
    V_{11} &= \frac{1}{4}\left( \eta_1\eta_2e^{2r_A} + (1-\eta_1)\eta_2 e^{2r_B} + 1-\eta_2(2\bar{n}_T + 1) \right)\\
    V_{22} &= \frac{1}{4}\left( \eta_1\eta_2e^{-2r_A} + (1-\eta_1)\eta_2 e^{-2r_B} + 1-\eta_2(2\bar{n}_T + 1) \right)\\
    V_{12}&=V_{21}=0.
\end{align}

Additionally, if Alice is constrained to mean photon number $\bar{n}_A$ and when $r_A \neq 0$, then she can only use $\bar{n}_\alpha = \bar{n}_A - \frac{1}{2}\cosh{(2r_A)} + \frac{1}{2} = \bar{n}_A - X_{1} - X_{2} + \frac{1}{2}$ photons for modulating the mean of her Gaussian input state with the rest of the photons reserved for squeezing. Similarly, for Bob: $\bar{n}_\beta = \bar{n}_B - Y_{1} - Y_{2} + \frac{1}{2}$. The fractional signal mean photon numbers received by Charlie from Alice and Bob are
\begin{align}
&N_C^A = \eta_1\eta_2 \bar{n}_\alpha = \eta_1\eta_2\left(\bar{n}_A - X_1 - X_2 + \frac{1}{2}\right)  \label{eq:nca} \\
&N_C^B = (1-\eta_1)\eta_2 \left(\bar{n}_B - Y_1 - Y_2 + \frac{1}{2}\right). \label{eq:ncb}
\end{align}

\subsection{Coherent Receivers}
 It is useful to examine coherent receivers when considering channel capacity for Gaussian states as they are commonly used in practice \cite[Section II.E]{weedbrook12gaussianQIrmp}. Coherent receivers are either homodyne or heterodyne. Homodyne detection measures a single quadrature, and is described by the positive operator-valued measure (POVM)  $\hat{\Pi}_{q}=\left| q\right\rangle\left\langle q\right|, \text { for } q \in \mathcal{R}$, where $| q\rangle$ are eigenstates of the real quadrature component $\hat{q}=\operatorname{Re}(\hat{a})$ of annihilation operator $\hat{a}$.

If Alice and Bob both transmit squeezed states with means $\bm{\bar{\alpha}} = [\bar{\alpha}_1 ~ \bar{\alpha}_2]^\top$ and $\bm{\bar{\beta}}  = [\bar{\beta}_1 ~ \bar{\beta}_2]^\top$  the quadrature mean and variance at Charlie are $\bar{x}_1 =\sqrt{\eta_1\eta_2}\bar{\alpha}_1 + \sqrt{(1-\eta_1)\eta_2}\bar{\beta}_1$ and $V_{1} = \frac{1}{4}(\eta_1\eta_2e^{-2r_A} + (1-\eta_1)\eta_2e^{-2r_B} + (1-\eta_2)(2\bar{n}_T + 1))$ with a similar expression for $V_2$. Homodyne detection measurement outcomes are governed by the Gaussian distribution conditioned on the quadrature mean $\bar{x}_1$ of Charlie's state
\begin{align}
p\left(q | \bar{x}_1 \right) =\frac{1}{\sqrt{2 \pi V_1}} \exp \left(-\frac{\left(q- \bar{x}_1\right)^{2}}{2V_1}\right).
\end{align}

Since the choice of homodyne receiver induces an additive white Gaussian noise (AWGN) channel, its capacity is maximized for a Gaussian input. The sum-rate  capacity \eqref{eq:sum-rate} of squeezed state homodyne detection for the Alice-to-Charlie channel in Fig.~\ref{fig:model} is \cite[eq. 7.51]{guha04mastersthesis}
\begin{align}
    C_{\rm{hom}} = \frac{1}{2}\log{\left(1+\frac{4(\bar{n}_\alpha + \frac{(1-\eta_1)}{\eta_1}\bar{n}_\beta)}{ e^{2r_A}+\frac{(1-\eta_1)}{\eta_1}e^{2r_B}+\frac{1-\eta_2}{\eta_1\eta_2}(1+2\bar{n}_T)}\right)}. \label{eq:homodyne}
\end{align}
Note that setting $r_A=r_B=0$ yields the coherent-state homodyne detection capacity.

One can describe heterodyne detection as homodyne detection along both quadratures. The heterodyne detection capacity derivation follows similarly to that of homodyne described above with a POVM $\hat{\Pi}_{a}=\frac{1}{\pi}|a\rangle\langle a|$ where $|a\rangle$ are eigenstates of $\hat{a}$. It is maximized when coherent states are employed ($r_A = r_B = 0$). The sum-rate capacity of coherent-state heterodyne detection is \cite[eq. 7.43]{guha04mastersthesis}
\begin{align}
    C_{\rm{het}} = \log{\left(1+\frac{\eta_1\bar{n}_A+(1-\eta_1)\bar{n}_B}{1+\frac{1-\eta_2}{\eta_2}(1+2\bar{n}_T)}\right)}, \label{eq:heterodyne}
\end{align}
and for individual user capacity heterodyne and homodyne detection:
\begin{align}
   C_{\rm{hetA}} = \left.C_{\rm{het}}\right|_{\bar{n}_{B}=0}, \quad C_{\rm{hetB}} &= \left.C_{\rm{het}}\right|_{\bar{n}_{A}=0} \label{eq:heterodyne_singlerate} \\
    C_{\rm{homA}} = \left.C_{\rm{hom}}\right|_{\bar{n}_{B}=0}, \quad C_{\rm{homB}} &= \left.C_{\rm{hom}}\right|_{\bar{n}_{A}=0}. \label{eq:homodyne_singlerate} 
\end{align}

\subsection{Quantum Gaussian MAC}\label{sec:gaussianmaccapacity}
Yen and Shapiro extended the Holevo-Sohma-Hirota classical capacity of quantum point-to-point Gaussian channels for squeezed states \cite{holevo_capacity_1999} to two users \cite{yen_multiple-access_2005,yen05phdthesis}. They found that the maximum rates of the channel are given by a piecewise function, which we restate here using the following three functions:

\begin{align}
 G_{11}(N, \bm{V}) & = g\left(V_{1}+V_{2}+N-\frac{1}{2}\right), \label{eq:G11} \\
 G_{12}(N, \bm{V})) & = g\left(2\left[ -\left(\sqrt{\left(\frac{V_{1}-V_{2}}{2}\right)^2+V_{12}^{2}}-\frac{N}{2}\right)^{2} 
 \right. \right. \notag
 \\ &\phantom{=g\left(2\left[\vphantom{\left(\sqrt{\frac{V_1}{V_2}^2}\right)^2}\right.\right.}+\left(\frac{V_{1}+V_{2}+N}{2}\right)^2 \left.\vphantom{\left[\sqrt{\frac{1}{2}}\right]^2} \right]^{\frac{1}{2}} -\left. \vphantom{\left[\left(\sqrt{\frac{1}{2}}\right)^2\right]^2} \frac{1}{2}\right), \label{eq:G12} \\
 G_{2}(\bm{V}) & = g\left(2|\bm{V}|^{1 / 2}-\frac{1}{2}\right) \label{eq:G2}
\end{align}
where $g(\cdot)$ is defined as 
\begin{align}
g(x) \equiv (1+x)\log{(1+x)} - x\log{x} \label{eq:g}.
\end{align} 
The sum rate for the MAC is
\begin{align}
    R_{\rm maxAB}=\left\{\def\arraycolsep{4pt}\begin{array}{ll}R_{\rm maxAB1}&\mbox{if   $N_C^{A+B} \geq\sqrt{\left(V_{1}-V_{2}\right)^{2}+4 V_{12}^{2}}$} \\
R_{\rm maxAB2} &\mbox{if   $N_C^{A+B} <\sqrt{\left(V_{1}-V_{2}\right)^{2}+4 V_{12}^{2}}$}\end{array}\right. \label{eq:rmaxab}
\end{align}
where $N_C^{A+B} = N_C^{A} + N_C^{B}$, $R_{\rm{maxAB1}} = G_{11}(N_C^{A+B}, \bm{V}) - G_{2}(\bm{V}$), and $R_{\rm{maxAB2}} = G_{12}(N_C^{A+B}, \bm{V}) - G_{2}(\bm{V})$. 
The individual rate is
\begin{align}
    R_{\rm maxA}=\left\{\def\arraycolsep{4pt}\begin{array}{ll}R_{\rm maxA1} & \mbox{if  $ N_C^{A} \geq\sqrt{\left(V_{1}-V_{2}\right)^{2}+4 V_{12}^{2}}$} \\ 
R_{\rm maxA2} & \mbox{if   $N_C^{A} <\sqrt{\left(V_{1}-V_{2}\right)^{2}+4 V_{12}^{2}}$}\end{array}\right. \label{eq:rmaxa}
\end{align}
where $R_{\rm{maxA1}} = G_{11}(N_C^{A}, \bm{V}) - G_{2}(\bm{V})$, and $R_{\rm{maxA2}} = G_{12}(N_C^{A}, \bm{V}) - G_{2}(\bm{V})$. An expression for $R_{\rm{maxB}}$ is obtained by swapping $N_C^A$ with $N_C^B$. For the states we are interested in, $V_{12}=0$ and the threshold
in \eqref{eq:rmaxa} evaluates to $N_C^A \lessgtr |V_1 - V_2|$. A similar condition also holds for $\eqref{eq:rmaxab}$.

\subsection{Coherent-State Gaussian Random Code} \label{sec:code}
Consider input with block size of $M$ bits and codewords of length $n$. Then a channel code's rate is $R=M/n$ bits/symbol. Let $\hat{\rho}(\alpha)$ be a coherent state with mean $\alpha\in\mathbb{C}$. Then suppose the encoder generates $2^{nR}$ codewords for codebook  $\mathcal{C}=\{\boldsymbol{\otimes}_{m=1}^n\hat{\rho}_k(\alpha_m)\}_{k=1}^{2^{nR}}$ each according to $p(\boldsymbol{\otimes}_{m=1}^n\hat{\rho}_k(\alpha_m))=\Pi_{m=1}^np(\alpha_m)$, where $p(\alpha) = e^{-|\alpha|^2/\bar{n}}/(\pi \bar{n})$ is the circularly-symmetric complex Gaussian distribution, $\boldsymbol{\otimes}_{i=1}^n$ denotes the $n$-mode tensor product, and $\bar{n}$ is the mean photon number per symbol. Yen and Shapiro showed that such a random code combined with a joint measurement receiver maximizes the sum rate $R_{\rm{maxAB}}$ for the pure-loss bosonic MAC. Moreover, such encoding also was shown to be optimal for the sum rate for the generalized phase-insensitive bosonic MAC \cite{shi_computable_2021}. We employ this code in the next section.

\section{Thermal-Noise Lossy Bosonic Multiple Access Channel Capacity} 
\label{sec:main}
In this section we investigate the capacity for the thermal-noise lossy bosonic MAC and its asymptotic limits.

\subsection{Individual Rate Outer Bounds} \label{sec:outerbounds}
 It was conjectured in \cite{giovannetti_minimum_2004} and later proven in \cite{giovannetti_ultimate_2014} that for the point-to-point thermal-noise lossy bosonic channel, the capacity is reached with coherent-state encoding. For a general point-to-point lossy thermal-noise channel, given $\bar{x}$ as the mean photon number of the signal from the transmitter at the receiver and $\bar{y}$ as the mean photon number of the thermal noise at the receiver, the capacity is:
\begin{align}
    C(\bar{x}, \bar{y}) = g(\bar{x} + \bar{y}) - g(\bar{y}) \label{eq:singlebound}.
\end{align} 
Thus, the capacity of Alice-to-Charlie (point-to-point) channel when Bob inputs vacuum is:
\begin{align}
C_{\text{A}} = C(\eta_1 \eta_2  \bar{n}_A, (1-\eta_2)\bar{n}_T). \label{eq:point-to-point}
\end{align}

For our two-user thermal-noise MAC, the ultimate upper bound,  (ub) regardless of input state (Gaussian or otherwise) for the individual rates ignores the interference from the other user. This corresponds to a nonphysical receiver that can undo the beamsplitter between the two transmitters yielding
\begin{align}
    R_{\rm{maxA}}\leq R_{\rm{ubA}}=C(\eta_2 \bar{n}_A, (1-\eta_2)\bar{n}_T) \label{eq:outerbound_a} \\
    R_{\rm{maxB}}\leq R_{\rm{ubB}}=C(\eta_2 \bar{n}_B, (1-\eta_2)\bar{n}_T). \label{eq:outerbound_b}
\end{align}

\subsection{Asymptotics of Photon-Number Constraints} \label{subsec:asymptotics}
The sum-rate capacity is achieved through the use of coherent Gaussian states \cite{giovannetti_ultimate_2014, shi_computable_2021}, 
\begin{align}
    R_{A} + R_{B} \le C(\eta_2(\eta_1\bar{n}_A + (1-\eta_1)\bar{n}_B), (1-\eta_2)\bar{n}_{T}), \label{eq:sumratecapacity}
\end{align} 
where the RHS is equivalent to substituting coherent states in \eqref{eq:rmaxab} and $C(\bar{x}, \bar{y})$ is defined in \eqref{eq:singlebound}. 
Now, we evaluate the scaling of the individual rate for Alice with respect to the upper bounds in \eqref{eq:outerbound_a} in the case of asymptotically large input power noting that the same process applies for Bob and \eqref{eq:outerbound_b}.

\begin{lemma} \label{lemma:1}
A random code with coherent-state encoding with heterodyne detection achieves the individual rate capacity of the thermal-noise lossy bosonic MAC in the asymptotic limit of large transmitter input power with constant thermal noise. 
\end{lemma}
\begin{IEEEproof} For coherent-state inputs ($r_{A}=r_{B}=0$) at Alice and Bob and a heterodyne detector at the receiver, Charlie, the capacities $C_{\rm{hetA}}$ and $C_{\rm{hetB}}$ are given by \eqref{eq:heterodyne_singlerate}. Then
\begin{align}
\lim_{\bar{n}_{A}\to\infty}\frac{C_{\rm{hetA}}}{R_{\rm{ubA}}} =  \lim_{\bar{n}_{B}\to\infty}\frac{C_{\rm{hetB}}}{R_{\rm{ubB}}} = 1 \label{eq:lim_lemma_1}.
\end{align} 
Evaluation of the limits are performed using L'H\^{o}pital's rule and derived in Appendix~\ref{appendix:lemma1_limit}. Hence, coherent-state encoding with heterodyne detection yields the individual rate upper bounds as $\bar{n}_A \to \infty$ and $\bar{n}_B \to \infty$.
\end{IEEEproof}

For homodyne detection utilizing coherent states or squeezed states, the scaling of the individual rate for Alice with respect to the upper bound in \eqref{eq:outerbound_a} is evaluated as follows: 
\begin{align}
\label{eq:homodyne_nAinft}\lim_{\bar{n}_{A}\to\infty}\max_{r_{A}}\lim_{\bar{n}_{B}\to\infty}\max_{r_{B}}\frac{C_{\rm{hom}}}{R_{\rm ubA}}&=\frac{1}{2},
\end{align}
where the maximization over $r_{B}$ in the inner limit as $\bar{n}_{B}\to\infty$ puts all the energy available to Bob into squeezing, that is, optimal $r_{B}\to-\infty$. 
This contrasts the pure-loss channel result in \cite{yen_multiple-access_2005, yen05phdthesis} where \eqref{eq:homodyne_nAinft} evaluates to unity.

\begin{lemma} \label{lemma:2} In the limit of small photon number and constant thermal noise, utilizing a random code with coherent-state encoding at Alice and Bob and a joint detection receiver at Charlie achieves the capacity of the thermal-noise lossy bosonic MAC.
\end{lemma}
\begin{IEEEproof} We evaluate the three cases corresponding to the order in which photon numbers input by Alice and Bob decay to zero.

\emph{Case 1.} Bob's input photon number decays to zero first:
\begin{align}
    \lim_{\bar{n}_A \to 0} \lim_{\bar{n}_B \to 0} \frac{ R_{\rm{maxA}}}{C_{\text{A}}} = 1 \label{eq:lem2_lim1}
\end{align}
When Bob's photon number decays to zero first, he effectively has no photons for squeezing guaranteeing that $N_C^A \ge |V_1-V_2|$ and $R_{\rm{maxA}}=R_{\rm{maxA1}}$. Hence, the MAC reduces to a point-to-point channel in which Alice's signal photon number is attenuated by an $\eta_1\eta_2$ term and coherent-state encoding achieves the capacity $C_{\text{A}}$ in \eqref{eq:point-to-point}.  This demonstrates that \eqref{eq:outerbound_a} can be made tighter in the asymptotic limit of small input power. Limits in \eqref{eq:lem2_lim1} are evaluated by inspection.

\emph{Case 2.} Alice's input photon number decays to zero first:
\begin{align}
    \lim_{\bar{n}_B \to 0} \lim_{\bar{n}_A \to 0} \frac{ R_{\rm{maxA}}}{C_{\text{A}}} = 1 \label{eq:lem2_lim2}
\end{align}
We allow Bob to perform an arbitrary amount of squeezing, setting his squeezing parameter to $r_B=\sinh^{-1}(\sqrt{\bar{n}_B})$. If Alice's input photon number decays to zero first, $N_C^A < |V_1-V_2|$ and $R_{\rm{maxA}}=R_{\rm{maxA2}}$. The first limit ($\bar{n}_A \to 0$) is evaluated using a single application of L'H\^{o}pital's rule, yielding a simple expression that allows the second limit ($\bar{n}_B \to 0$) to be evaluated by inspection. Details are in Appendix \ref{appendix:lemma2_case2_limit}.

\emph{Case 3.} Alice and Bob's input photon numbers decay to zero simultaneously. Let $\bar{n}_A = a\bar{n}, \bar{n}_B = b\bar{n}$ with arbitrary constants $a,b > 0$. We show that
\begin{align}
    \lim_{\bar{n} \to 0} \frac{ R_{\rm{maxA}}}{C_{\text{A}}} = 1. \label{eq:lim_n_to_0}
\end{align}

First consider $N_C^A \ge |V_1-V_2|$ when $R_{\rm{maxA}} = R_{\rm{maxA1}}$. Then any squeezing at Alice is sub-optimal because
\begin{align}
    G_{11}(N_C^A, \bm{V}) & = g\left(\eta_1 \eta_2 a \bar{n} + \left(1-\eta_1\right)\eta_2b\bar{n} + (1-\eta_2)\bar{n}_T \right) \notag
\end{align} is a function of the total photon number $a\bar{n}$ at Alice and does not depend on squeezing parameter $r_A$. Furthermore $|\bm{V}|$ in \eqref{eq:G2} is minimized for coherent-state input at Alice. Therefore, since $g(\cdot)$ is monotonic, $R_{\rm{maxA1}}$ is maximized by setting $r_A=0$. When Bob and Alice both use coherent-state encoding:
\begin{align}
\lim_{\bar{n}\to 0} \frac{R_{\rm{maxA1}}|_{r_A=r_B=0}}{C_{\text{A}}} = 1 \label{eq:limit_a_co_b_co}.
\end{align}

To show that any squeezing at Bob’s transmitter cannot help, we first note that the only impact from Bob’s transmissions on  $R_{\rm maxA1}$ is through transmission of squeezed states. Suppose that Bob is allowed to squeeze arbitrarily. As $b$ is arbitrary, let Bob's squeezing parameter be $r_B=\sinh^{-1}(\sqrt{b \bar{n}})$, and Alice is transmitting an optimal coherent state ($r_A=0)$ as discussed previously. However, the constraint $N_C^{A} \geq |V_{1}-V_{2}|$ upper bounds $b$ as
\begin{align}
b_{\rm{maxA1}} \le \frac{\eta_1 - 1 + \sqrt{1+\eta_1(2-\eta_1(1-4a^2 \bar{n}^2))}}{2(1-\eta_1) \bar{n}}. \label{eq:bmaxa1}
\end{align}
Note that $b$ is dependent on $\bar{n}$. Then, to upper bound Bob's possible arbitrary squeezing, let $b=\kappa b_{\rm{maxA1}}$ where $\kappa$ is arbitray and $\kappa \in [0,1]$. Hence the limit involving $R_{\rm{maxA1}}$ becomes
\begin{align}
\lim_{\bar{n}\to 0} \frac{R_{\rm{maxA1}}|_{r_A=0,r_B=\sinh^{-1}(\sqrt{\kappa b_{\rm{maxA1}} \bar{n}})}}{C_{\text{A}}} = 1. \label{eq:limit_a_co_b_sq_max}
\end{align}
Evaluation of the limit involves an application of L'H\^{o}pital's rule and is shown in Appendix \ref{appendix:lemma2_case3_limit1}. Thus, squeezing does not help when $N_C^{A} \geq |V_{1}-V_{2}|$.

Now consider $N_C^{A} < |V_{1}-V_{2}|$ when $R_{\rm{maxA}} = R_{\rm{maxA2}}$. For Gaussian state inputs, $G_{12}(N_C^A, \bm{V})$ = $g(2\sqrt{V_1(\eta_1 \eta_2 p_A a \bar{n} + V_2)} - \frac{1}{2})$ for $V_1 > V_2$ (with $V_1$ and $V_2$ swapped when $V_2 > V_1$), where $p_A \in [0,1]$ is the fraction of photons Alice uses for displacement and $(1-p_A)$ the fraction used for squeezing. Additionally, for coherent-state encoding, $G_{11}(N_C^A, \bm{V}) = g(\eta_1 \eta_2 a \bar{n} + (1-\eta_2)\bar{n}_{T})$. Now
\begin{align}
\lim_{\bar{n}\to 0} \frac{G_{12}(N_C^A, \bm{V})}{G_{11}(N_C^A, \bm{V})|_{r_A=r_B=0}} = 1, \label{eq:gfunclimit1}
\end{align}
which is derived using meticulous expansion of the terms and taking their corresponding limit in Appendix \ref{appendix:lemma2_case3_limit2}. Thus, because the definitions of $R_{\rm{maxA1}}$ and $R_{\rm{maxA2}}$ include $G_2(\bm{V})$ terms  
\begin{align}
\label{eq:equationlimits}\lim_{\bar{n}\to 0} \frac{R_{\rm{maxA2}}}{R_{\rm{maxA1}}|_{r_A=r_B=0}} \le 1.
\end{align}

Squeezing by Alice or Bob increases $G_2(\bm{V})$ as $g(\cdot)$ is monotonic, and when $\bar{n}_T > 0$, $|\bm{V}|$ is an increasing function of $r_A$ and $r_B$. Thus, including $G_2(\bm{V})$ terms causes the numerator of \eqref{eq:gfunclimit1} to decay faster than the denominator. Therefore, any squeezing by Alice and Bob in $R_{\rm{maxA2}}$ is at best equivalent to the optimal rate defined by $R_{\rm{maxA1}}$ with $r_A=r_B=0$ in the asymptotic limit of small input power.

Employing Bob's maximum rate $R_{\rm{maxB}}$ and corresponding $C_{\text{A}} = C((1-\eta_1) \eta_2  \bar{n}_B, (1-\eta_2)\bar{n}_T)$ instead of $R_{\rm{maxA}}$  results in limits above also evaluating to $1$ as expected.
\end{IEEEproof} 

We note that when we replace $R_{\rm{maxA}}$ in \eqref{eq:lim_n_to_0} with coherent or squeezed homodyne or coherent heterodyne capacities, the limit evaluates to zero. Therefore achieving bosonic MAC capacity requires more complex receiver designs.

\subsection{Fixed Mean Photon-Number Constraints}\label{sec:fixed-mean-photon}
While utilization of coherent states are optimal at both asymptotically high and low input mean photon number, most practical systems operate in a finite mean photon number regime of fixed input power. There are two possible ways to fix the input power, via a global mean photon-number constraint that is spread across all transmitters, and a mean photon-number constraint for each of the individual transmitters.

We first examine the global mean photon number constraint $\bar{n}_S$, for both the transmitters. Let the fractional photon-number constraint for Alice be $\bar{n}_A = s\bar{n}_S$ and likewise for Bob, $\bar{n}_B = (1-s)\bar{n}_S$ where $s \in [0,1]$. As this constraint is global, we assume Alice and Bob can control $s$. Only the photons Bob uses for squeezing affect Alice's rate as they introduce additional noise in the channel. In this case, we can treat the channel as a point-to-point channel from Alice to Charlie with a mean photon-number constraint $\bar{n}_S$ such that $s\bar{n}_S$ photons are used for the signal and $(1-s)\bar{n}_S$ photons are used for squeezing by the transmitter. As mentioned in Sec.~\ref{sec:outerbounds}, coherent states are optimal in the point-to-point channel, which corresponds to Alice using all $\bar{n}_S$ photons for displacement, or when $s=1$. For the sum rate $R_{\rm{maxAB}}$, substituting  \eqref{eq:covariancerelationship}, \eqref{eq:nca}, and \eqref{eq:ncb} into $G_{11}(N, \bm{V})$ yields $G_{11}(N_C^{A+B}, \bm{V}) = g(\bar{n}_{T} + (\bar{n}_B - \bar{n}_{T} + (\bar{n}_A - \bar{n}_B)\eta_1)\eta_2)$, without squeezing parameter dependence. Similarly, $G_{12}(N_C^{A+B}, \bm{V})$ is strictly smaller than $G_{11}(N_C^{A+B}, \bm{V})$. Thus maximization of  $R_{\rm{maxAB}}$ reduces to the minimization of of $G_{2}(\bm{V})$, or $|\bm{V}|$, as $g(\cdot)$ is monotonic. The minimum occurs for coherent-state inputs ($r_A = r_B = 0$) and matches the results in \cite{yen_multiple-access_2005, yen05phdthesis, shi_computable_2021}.

\begin{figure}[htbp]
\centerline{\includegraphics[width=0.4\textwidth]{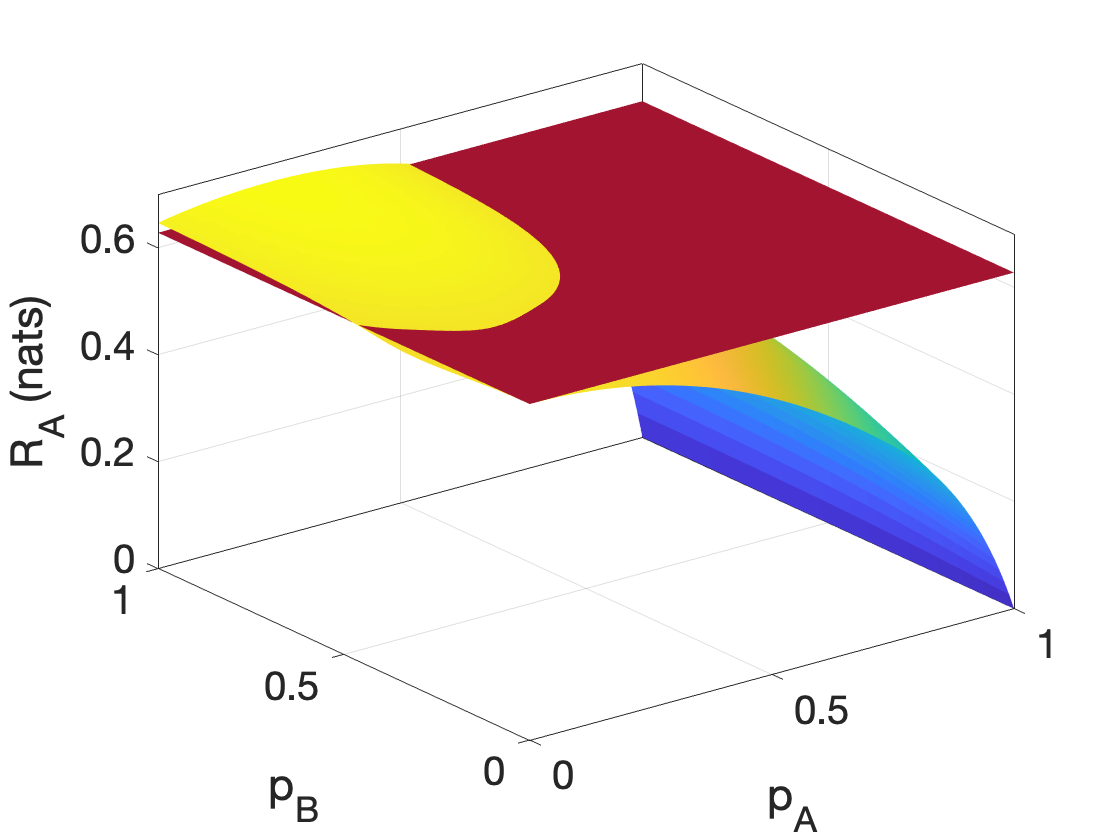}}
\caption{Individual rates generated for all possible squeezed states given input parameters $\bar{n}_A = 4, \bar{n}_B=8, \bar{n}_T=4, \eta_1 = 0.2, \eta_2 = 0.9$. $p_A= \sinh^2(r_A)/\bar{n}_A$ and $p_B= \sinh^2(r_B)/\bar{n}_B$ are fractions of the total photon number utilized for squeezing by Alice and Bob respectively. The dark red plane indicates the rate achieved utilizing coherent-state encoding ($p_A=p_B=0$) at both transmitters.}
\label{fig:squeezingh}
\end{figure}

\begin{figure}[htbp]
\centerline{\includegraphics[width=0.5\textwidth]{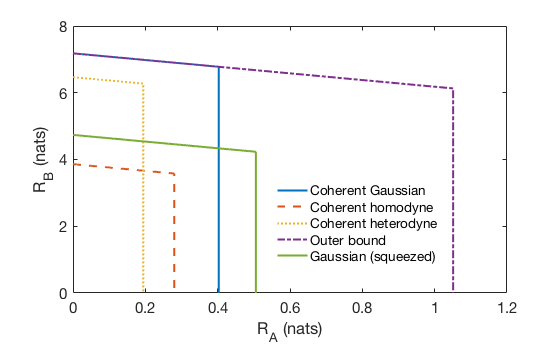}}
\caption{Capacity region given for rates \eqref{eq:rmaxab}, \eqref{eq:rmaxa} (and corresponding $R_{\rm{maxB}}$) for the use of coherent Gaussian states and joint detection, coherent homodyne, coherent heterodyne detection, Gaussian squeezed states, and the outer bound defined in \eqref{eq:outerbound_a}. Here, $\bar{n}_A=1, \bar{n}_B=1000, \bar{n}_T = 1, \eta_1 = 0.25, \eta_2 = 0.9$, and in the case of the squeezed state region, $r_A = 0, r_B = 3$.}
\label{fig:capacityplot}
\end{figure}

On the other hand, if the mean photon-number constraints are for the individual transmitters i.e. the transmitters cannot share their power, coherent-state encoding does not necessarily maximize the individual rates for the lossy thermal-noise bosonic MAC. While squeezing does not encode information directly, an example showing that squeezed states outperform encoding using only coherent states is shown in Fig.~\ref{fig:squeezingh}. The non-linear optimization of squeezing parameters is analytically intractable. However, given the parameters of a channel, a numeric search quickly shows whether squeezing helps and the corresponding values for the squeezing parameters. In Fig.~\ref{fig:capacityplot} we plot the capacity rate region for coherent homodyne, coherent heterodyne, coherent Gaussian encoding (with joint detection), an example of using squeezed states, and the individual rate outer bound in \eqref{eq:outerbound_a}. In this example, utilizing squeezed states pushes out $R_A$ beyond the coherent Gaussian encoding envelope at the expense of reduced $R_B$.

\section{Discussion and Conclusion}
\label{sec:discussion}

When there is a total mean photon-number constraint on the system, Alice and Bob can do no better than the capacity of coherent-state encoding, allocating all of the power to Alice or Bob, depending on the transmission goal. For individual mean photon-number constraints, or when Alice and Bob lack control over total power allocation, Figs.~\ref{fig:squeezingh} and \ref{fig:capacityplot} show that squeezing can improve the individual rates, although an intuitive explanation for this remains elusive. Characterizing the benefit of such "local" quantum enhancement at the transmitters is a compelling direction for future work. It is possible that the use of non-Gaussian states can reach the outer bound and further exploration of these states remains an open problem as well.

In Lemmas~\ref{lemma:1}-\ref{lemma:2} we prove that in asymptotic regimes of both very high or very low SNR, coherent-state encoding is optimal at the transmitters. The fact that coherent states are capacity-achieving for asymptotically low mean photon number lays the foundation for performing covert communication analysis for the thermal-noise bosonic MAC, and is a logical next step. However, unlike the high SNR regime, neither homodyne nor heterodyne detection are capacity-achieving, and other physically realizable receiver designs should be explored.

\section*{Acknowledgment}

The authors thank Saikat Guha for helpful and insightful discussion.

\appendix

The limits in Section \ref{sec:main} are derived utilizing the properties found in most undergraduate calculus textbooks.  We list them here for reference. For arbitrary real-valued functions $f(x), h(x)$ and constant $c$:
\begin{align}
    &\lim_{x \to a} cf(x) = c \lim_{x \to a}f(x) \label{eq:lim_cf} \\
    &\lim_{x \to a} f(x)h(x) = \lim_{x \to a}f(x)\lim_{x \to a} h(x) \label{eq:lim_f_times_h} \\
    &\lim_{x \to a} \frac{f(x)}{h(x)} = \frac{\lim_{x \to a}f(x)}{\lim_{x \to a}  h(x)}, \text{ if } \lim_{x \to a}  h(x) \neq 0 \label{eq:lim_f_divided_h} \\
    &\lim_{x \to a} f(x) \pm h(x) = \lim_{x \to a}f(x) \pm \lim_{x \to a} h(x) \label{eq:lim_f_pm_h} \\
    &\lim_{x \to a} f(h(x)) = f(\lim_{x \to a}h(x)), \label{eq:lim_f_of_h}
\end{align}
where the limits must exist and \eqref{eq:lim_f_of_h} is applicable only when $\lim_{x \to a}h(x)$ is in the domain of $f$. \eqref{eq:lim_f_of_h} allows for L'H\^{o}pital's rule to be applied as well as two useful instances in our limits: $\lim_{x \to a} \log(f(x)) = \log(\lim_{x \to a} f(x))$ and $\lim_{x \to a} \sqrt{f(x)} = \sqrt{\lim_{x \to a} f(x)}$ when $f(a) > 0$. 

We also use the following on hyperbolic trigonometric identities:
\begin{align}
    \cosh{x} = \frac{1}{2}(e^{x} + e^{-x}) \\ 
    \sinh{x} = \frac{1}{2}(e^{x} - e^{-x}) \\ 
    e^{\sinh^{-1}\sqrt{x}} = \sqrt{x}+\sqrt{x+1} \label{eq:htrig3}
\end{align}

We denote the first and second derivatives of the function $f(x)$ by $f^\prime(x)$ and $f^{\prime \prime}(x)$.

\subsection{Evaluation of the Limit in \eqref{eq:lim_lemma_1} in the Proof of Lemma~\ref{lemma:1}}\label{appendix:lemma1_limit}
Taking the forms of $C_{\rm{hetA}}$ and $R_{\rm{ubA}}$ in \eqref{eq:heterodyne_singlerate}, \eqref{eq:outerbound_a} respectively and applying \eqref{eq:lim_f_divided_h} to \eqref{eq:lim_lemma_1} yields the indeterminate form of $\infty/\infty$. The first application of L'H\^{o}pital's rule with derivatives taken with respect to $\bar{n}_A$ yields:
\begin{align}
&\lim_{\bar{n}_{A}\to\infty}\frac{C^\prime_{\rm{hetA}}}{R^\prime_{\rm{ubA}}} \\ 
&=\lim_{\bar{n}_{A}\to\infty} \frac{\frac{\eta_1\eta_2}{\eta_2+(1-\eta_2)(1+2\bar{n}_T)+\eta_1\eta_2\bar{n}_A}}{\gamma} \label{eq:lhopital_het_limit}
\end{align}
Where $\gamma=\eta_2(\log(1+\eta_2\bar{n}_A+(1-\eta_2)\bar{n}_T)-\log(\eta_2\bar{n}_A+(1-\eta_2)\bar{n}_T)$. Applying \eqref{eq:lim_f_divided_h} to \eqref{eq:lhopital_het_limit} yields the indeterminate form of $0/0$ and hence a second application of L'H\^{o}pital's rule yields a simplified form 
\begin{align}
&\lim_{\bar{n}_{A}\to\infty}\frac{C^{\prime\prime}_{\rm{hetA}}}{R^{\prime\prime}_{\rm{ubA}}}\\
&=\lim_{\bar{n}_{A}\to\infty} \frac{k_2}{(1+\eta_1\eta_2\bar{n}_A+2(1-\eta_2)\bar{n}_T)^2} = 1.
\end{align}
Where $k_2=\eta_1^2(\eta_2(\bar{n}_A-\bar{n}_T)+\bar{n}_T)(1+\eta_2(\bar{n}_A-\bar{n}_T)+\bar{n}_T)$ and the limit is evaluated through the application of the properties of limits described earlier noting that the highest order terms in both the numerator and denominator are equivalent to $\eta_1^2\eta_2^2\bar{n}_A^2$.

\subsection{Limit Evaluations for the Proof of Lemma~\ref{lemma:2}}\label{appendix:lemma2_limit}

\subsubsection{Evaluation of the Limit in \eqref{eq:lem2_lim2} }\label{appendix:lemma2_case2_limit}
Taking the functional forms of the numerator and denominator defined in \eqref{eq:rmaxa} and \eqref{eq:point-to-point}, and taking their limits as $\bar{n}_A \to 0$ gives the indefinite form of $0/0$, hence we can apply L'H\^{o}pital's rule. 

First note that $C_{\text{A}}$ is defined in terms of $C(\bar{x},\bar{y})$. The derivative with respect to $\bar{x}$ of $C(\bar{x},\bar{y})$ is
\begin{align}
C^\prime(\bar{x},\bar{y})=\log(1+\bar{x}+\bar{y}) - \log(\bar{x}+\bar{y}).
\end{align}
Then we can define the derivative of $C_{\text{A}}$ with respect to $\bar{n}_A$ as
\begin{align}
   C_{\text{A}}^\prime = \eta_1\eta_2C^\prime(\eta_1 \eta_2  \bar{n}_A,(1-\eta_2)\bar{n}_T), \label{eq:caprime} 
\end{align}
noting the additional constant factor of $\eta_1\eta_2$.

Next, recalling $R_{\rm{maxA2}} = G_{12}(N_C^A, \bm{V}) - G_2(\bm{V})$, we simplify $G_{12}(N_C^A, \bm{V})$ when Alice is using coherent states:
\begin{align}
G_{12}(N_C^A, \bm{V}) = g\left(2\sqrt{V_1(\eta_1 \eta_2 \bar{n}_A + V_2)} - \frac{1}{2}\right).
\end{align}
$G_2(\bm{V})$ is the same as in \eqref{eq:G2}, noting no dependence on $\bar{n}_A$ due to Alice's use of coherent states. Then $R_{\rm{maxA2}}^\prime$ with respect to $\bar{n}_A$ is
\begin{align}
&R_{\rm{maxA2}}^\prime  = G_{12}^\prime(N_C^A, \bm{V}) + G_2^\prime(\bm{V})  + 0 \\
    &= \frac{\eta_1\eta_2 V_1}{\sqrt{V_1(\eta_1\eta_2\bar{n}_A+ V_2)}} \\
    &\phantom{=}\times\log\left(\frac{1+4\sqrt{V_1(\eta_1\eta_2\bar{n}_A+ V_2)}}{-1+4\sqrt{V_1(\eta_1\eta_2\bar{n}_A+ V_2)}}\right).
\end{align}

Through L'H\^{o}pital's rule 
\begin{align}
&\lim_{\bar{n}_B \to 0} \lim_{\bar{n}_A \to 0} \frac{ R^\prime_{\rm{maxA2}}}{C^\prime_{A}} \\ 
&=\lim_{\bar{n}_B \to 0} \lim_{\bar{n}_A \to 0} \frac{ G_{12}^\prime(N_C^A, \bm{V})}{\eta_1\eta_2C^\prime(\eta_1 \eta_2  \bar{n}_A,(1-\eta_2)\bar{n}_T)} \\
&=\lim_{\bar{n}_B \to 0} \frac{\frac{V_1}{\sqrt{V_1 V_2}}\log\left(\frac{1+4\sqrt{V_1 V_2}}{-1+4\sqrt{V_1 V_2}}\right)}{\log(1+\frac{1}{(1-\eta_2)\bar{n}_T})}, \label{eq:lhop_limmb_to_zero}
\end{align}
where the limit of $\bar{n}_A$ can be evaluated by inspection and utilizing the properties of limits defined previously.
In order to perform the limit of $\bar{n}_B \to 0$, we first expand $V_1$ and $V_2$:

\begin{align}
V_1 &=  \frac{1}{4}(1-\eta_2)(2\bar{n}_T + 1) \notag \\ 
&\phantom{=}+ \frac{1}{4} (1-\eta_1)\eta_2e^{\sinh^{-1}(\sqrt{\bar{n}_B})} + \frac{1}{4}\eta_1\eta_2 \label{eq:appendix_v1_bsq}\\
V_2 &= \frac{1}{4}(1-\eta_2)(2\bar{n}_T + 1) \notag \\ 
&\phantom{=}+ \frac{1}{4}(1-\eta_1)\eta_2e^{-\sinh^{-1}(\sqrt{\bar{n}_B})} + \frac{1}{4}\eta_1\eta_2. \label{eq:appendix_v2_bsq}
\end{align}
Using \eqref{eq:lim_f_of_h} and \eqref{eq:lim_f_times_h} we only need to perform the limit for $V_1$ and $V_2$: $\lim_{\bar{n}_B \to 0} V_1 = \lim_{\bar{n}_B \to 0} V_2 = \frac{1}{4}( (1-\eta_2)(2\bar{n}_T + 1) + \eta_1\eta_2)$, and hence \eqref{eq:lhop_limmb_to_zero} evaluates to unity.

\subsubsection{Evaluation of the Limit in \eqref{eq:limit_a_co_b_sq_max}}\label{appendix:lemma2_case3_limit1} Applying the properties of limits to the limit in \eqref{eq:limit_a_co_b_sq_max} yields an indeterminate form of $0/0$ and we rely on L'H\^{o}pital's rule. The derivative with respect to $\bar{n}$ in the denominator, $C_A^\prime$, has no $b$ dependence and is given by:
\begin{align}
    C_A^\prime &= a\eta_1\eta_2 [ \log(1+\eta_1 \eta_2 a\bar{n} (1-\eta_2)\bar{n}_T) \label{eq:caprime_expanded}\\
    &\phantom{=} - \log(\eta_1 \eta_2  a\bar{n}+ (1-\eta_2)\bar{n}_T)].\notag
\end{align}
This limit can be performed by inspection,
\begin{align}
\lim_{\bar{n}\to0}C_A^\prime = a\eta_1\eta_2 [ \log(1+ (1-\eta_2)\bar{n}_T) - \log((1-\eta_2)\bar{n}_T)]. \label{eq:limit_caprime}
\end{align}
 The derivative of the numerator, $R^\prime_{\rm{maxA1}}|_{r_A=0,r_B=\sinh^{-1}(\sqrt{\kappa b_{\rm{maxA1}} \bar{n}})}$, can be split into two parts as $R^\prime_{\rm{maxA1}}=G_{11}^\prime(N_C^A, \bm{V})-G_2^\prime(\bm{V})$. Taking the derivative of the two terms yields
 \begin{align}
     G_{11}^\prime(N_C^A, \bm{V}) &= \frac{\theta_2}{\theta_1}(\log(1+\theta_3)-\log(\theta_3)) \label{eq:g11_prime_thetas} \\
     G_2^\prime(\bm{V}) &= \frac{\theta_5}{\theta_1\theta_4}(\log(1+\theta_4)-\log(-1+\theta_4))\label{eq:g2_prime_thetas},
\end{align}
where
\begin{align}
\theta_1 &= [1 - \eta_1 (2 - \eta_1 - 4 a^2 \eta_1 \bar{n}^2)]^\frac{1}{2} \\
     \theta_2 &= a \eta_1\eta_2(2a\eta_1 \kappa \bar{n}+\theta_1) \\
     \theta_3 &= a \eta_1\eta_2\bar{n}-\frac{1}{2}\eta_2\kappa(1-\eta_1-\theta_1)+(1-\eta_1)\bar{n}_T \\
     \theta_4 &= \left[ 2 \eta_2 \kappa (\eta_1 + \theta_1 - 1) (1 - \eta_2 (1 - \eta_1 + 2 \bar{n}_T) \right. \\ 
     &\phantom{=[}\left.+ \bar{n}_T) + (1 + 2 (1 - \eta_2) \bar{n}_T)^2  \right]^\frac{1}{2}\\
     \theta_5 &= 2 a^2 \eta_1^2 \eta_2 \kappa \bar{n} (1 - \eta_2 (1 - \eta_1 + 2 \bar{n}_T) + 2 \bar{n}_T).
\end{align}

Using \eqref{eq:lim_f_pm_h} we can additionally split the limit into two parts. In order to take the limit as $\bar{n}\to0$ of \eqref{eq:g2_prime_thetas}, note that the denominator and log terms are strictly non-zero as $\theta_4$ and $\theta_1$ only have partial $\bar{n}$ dependence, additionally $\theta_5$ has an $\bar{n}$ factor which goes to $0$, hence
\begin{align}
    \lim_{\bar{n}\to0} G_2^\prime(\bm{V}) = 0.
\end{align}

To find the limit as $\bar{n}\to0$ of \eqref{eq:g11_prime_thetas} we apply \eqref{eq:lim_f_divided_h} and take the limits of the numerator and denominator. While the equations are lengthy, the limits can be performed through inspection and use of eqs.~\eqref{eq:lim_cf}-\eqref{eq:lim_f_of_h}. After the limit is evaluated through inspection additional algebraic steps can be performed to simplify as
\begin{align}
    \lim_{\bar{n}\to0}  G_{11}^\prime(N_C^A, \bm{V}) &= a\eta_1\eta_2 [ \log(1+ (1-\eta_2)\bar{n}_T) \notag \\ 
    &\phantom{=a\eta_1\eta_2 [ }- \log((1-\eta_2)\bar{n}_T)].
\end{align}
This matches the results of the limit in \eqref{eq:limit_caprime} and hence the limit in \eqref{eq:limit_a_co_b_sq_max} evaluates to unity.

\subsubsection{Evaluation of the Limit in \eqref{eq:gfunclimit1}}\label{appendix:lemma2_case3_limit2}
For an arbitrary amount of squeezing at Bob, as in the previous limits, $r_B=\sinh^{-1}(\sqrt{b \bar{n}})$. We define $p_A$ to be the fraction of photons Alice uses for her signal and $(1-p_A)$ are used for squeezing. Therefore $r_A=\sinh^{-1}(\sqrt{a (1-p_A) \bar{n}})$. Here, 
\begin{align}
V_1 =& \frac{1}{4}( (1-\eta_2)(2\bar{n}_T + 1)+ (1-\eta_1)\eta_2e^{\sinh^{-1}(\sqrt{b \bar{n}})} \notag
\\ &+ \eta_1\eta_2e^{\sinh^{-1}(\sqrt{a (1-p_A) \bar{n}})} ) \label{eq:appendix_v1_absq}\\
V_2 =& \frac{1}{4}( (1-\eta_2)(2\bar{n}_T + 1)+ (1-\eta_1)\eta_2e^{-\sinh^{-1}(\sqrt{b \bar{n}})} \notag
\\ &+ \eta_1\eta_2e^{\sinh^{-1}(-\sqrt{a (1-p_A) \bar{n}})} ). \label{eq:appendix_v2_absq}
\end{align}
Note that similar to Appendix \ref{appendix:lemma1_limit}, $G_{12}(N_C^A, \bm{V})$ = $g(2\sqrt{V_1(\eta_1 \eta_2 p_A a \bar{n} + V_2)} - \frac{1}{2})$.

Additionally,
\begin{align}
    G_{11}(N_C^A, \bm{V})|_{r_A=r_B=0} = g(\eta_1\eta_2a\bar{n} + (1-\eta_2)\bar{n}_T).
\end{align}
The limit can be evaluated using \eqref{eq:lim_f_divided_h}. Note that the limit for $G_{12}(N_C^A, \bm{V})$ only needs to be evaluated for \eqref{eq:appendix_v1_absq} and \eqref{eq:appendix_v2_absq} due to \eqref{eq:lim_f_of_h}. By inspection and appropriate substitutions, 
\begin{align}
& \lim_{\bar{n}\to 0} \frac{G_{12}(N_C^A, \bm{V})}{G_{11}(N_C^A, \bm{V})|_{r_A=r_B=0}} \\
& = \frac{\lim_{\bar{n}\to 0} g(2\sqrt{V_1(\eta_1 \eta_2 p_A a \bar{n} + V_2)} - \frac{1}{2})}{\lim_{\bar{n}\to 0} g(\eta_1\eta_2a\bar{n} + (1-\eta_2)\bar{n}_T)} = 1.
\end{align}

\bibliographystyle{IEEEtran}
\bibliography{./papers.bib}

\end{document}